\documentclass[twocolumn]{aastex701} 

\usepackage{amsmath,amssymb}
\usepackage{overpic}

\begin{document}

\title{EP241113a: dissipative photospheric emission from a dirty fireball}

\author[0000-0002-0170-0741]{Cui-Yuan Dai}
\affiliation{School of Astronomy and Space Science, Nanjing University, Nanjing 210023, China; xywang@nju.edu.cn}
\affiliation{Key Laboratory of Modern Astronomy and Astrophysics (Nanjing University), Ministry of Education, Nanjing
210023, China}
\email[]{cydai@smail.nju.edu.cn}

\author[0000-0002-5881-335X]{Xiang-Yu Wang}
\affiliation{School of Astronomy and Space Science, Nanjing University, Nanjing 210023, China; xywang@nju.edu.cn}
\affiliation{Key Laboratory of Modern Astronomy and Astrophysics (Nanjing University), Ministry of Education, Nanjing
210023, China}
\email[]{xywang@nju.edu.cn}  

\author[0000-0002-9725-2524]{Bing Zhang}
\affiliation{The Hong Kong Institute for Astronomy and Astrophysics, the University of Hong Kong, Pokfulam Road, Hong Kong, China.}
\affiliation{Department of Physics, University of Hong Kong, Pokfulam Road, Hong Kong 999077, China}
\affiliation{Nevada Center for Astrophysics and Department of Physics and Astronomy, University of Nevada, Las Vegas, NV 89154, USA.}
\email[]{bzhang1@hku.hk}  

\begin{abstract}
EP241113a, a soft X-ray transient detected by the Einstein Probe (EP), has an isotropic-equivalent energy ($E_{\gamma,\rm{iso}}\sim 10^{51}{\, \rm erg}$) comparable to classical gamma-ray bursts (GRBs) but an exceptionally low peak energy ($E_{\rm{p}} \lesssim 1 \, {\rm keV}$), placing it off the canonical $E_{\rm{p}}$--$E_{\gamma,\rm{iso}}$ (Amati) relation at $>3\sigma$ confidence. The afterglow of EP241113a is also unusual, with an extremely low plateau luminosity and no jet break up to the latest observation. Such properties have been interpreted as arising from an energetic ``dirty fireball,'' i.e., a relativistic jet with energy comparable to classical GRBs but much higher baryon loading, leading to a bulk Lorentz factor of $\Gamma\sim 20$. In this {\it Letter}, we propose that low $E_{\rm{p}}$ arises from dissipative photospheric emission in such a low-$\Gamma$ jet. As $\Gamma$ decreases, the photosphere shifts outward while the internal shock radius moves inward, causing dissipation to occur well below the photosphere and making the prompt emission photosphere-dominated. Meanwhile, the thermalization radius also moves outward, reducing the comoving radiation temperature where the spectral peak is established. Combined with weaker Lorentz boosting, these effects naturally shift the observed $E_{\rm{p}}$ into the soft X-ray band. We further suggest that such a low-$\Gamma$ jet may be powered by neutrino--antineutrino ($\nu\bar{\nu}$) annihilation in a hyperaccreting disk formed during the core collapse of a massive star. Compared with the Blandford--Znajek process, neutrino annihilation can produce a dirtier jet through baryon entrainment from a neutrino-driven wind. EP241113a-like events therefore offer new insight into the dissipation physics and launching mechanisms of jets from collapsing massive stars.

\end{abstract}

\section{Introduction}
\label{sec:intro}

Classical gamma-ray bursts (GRBs) are widely interpreted as ultra-relativistic, collimated outflows launched during the collapse of a massive star or the merger of compact objects, with low baryon loading and typical bulk Lorentz factors $\Gamma \gtrsim 100$ \citep[e.g.,][]{Eichler1989,Narayan1992,Woosley1993,MacFadyenWoosley1999,Piran1999,Meszaros2006,KumarZhang2015}. Long GRBs follow a broad correlation between the rest-frame spectral peak energy $E_{\rm p}$ and the isotropic-equivalent radiated energy $E_{\gamma,\rm iso}$, approximately $E_{\rm p}\propto E_{\gamma,\rm iso}^{1/2}$, known as the Amati relation \citep{Amati2002,Amati2006}. This trend appears to extend from classical long GRBs to X-ray flashes (XRFs), spanning peak energies from the MeV band down to $\sim$keV \citep[e.g.,][]{Barraud2003,Sakamoto2005,Sakamoto2008}.

EP241113a, recently discovered by \emph{Einstein Probe} (EP), is a significant outlier from the Amati relation, deviating from it at a confidence level of $\gtrsim 3\sigma$. Its prompt emission has an isotropic-equivalent radiated energy of $E_{\gamma,\rm iso}\sim10^{51}\,{\rm erg}$, comparable to that of classical GRBs, yet an exceptionally soft spectrum with $E_{\rm p}<{\rm a\ few\ keV}$ and no detected $\gamma$-ray counterpart despite simultaneous coverage by \emph{Fermi}/GBM and \emph{Swift}/BAT. Its afterglow properties are also unusual. In particular, the X-ray plateau luminosity is much lower than that of  GRBs that have a comparable isotropic-equivalent energy. The temporal behavior of the plateau can be interpreted as the coasting phase of a jet propagating into a stellar-wind environment. The low-luminosity plateau then implies a low Lorentz factor of $\Gamma\sim20$ \citep{Dai2026}. The absence of a jet break out to $\sim30$~days further suggests a relatively wide outflow. Together, these properties point to a long-anticipated ``dirty fireball'': a baryon-loaded jet with a moderate Lorentz factor but a substantial energy budget \citep[e.g.,][]{Paczynski1998,Dermer1999,Huang2002,Rhoads2003,ZhangWoosley2003}.

Although dirty fireballs have been discussed for decades, observational evidence has remained limited and largely indirect. A few orphan optical afterglows have been proposed as possible examples of baryon-rich outflows \citep{Cenko2013,Ho2022}, but their prompt emission has not been securely identified. EP241113a therefore provides a rare opportunity to probe the prompt emission of a dirty fireball, particularly the radiation mechanism of such an energetic yet unusually soft transient and the origin of its low-$\Gamma$ jet.

In this work, we argue that the prompt emission of EP241113a is naturally explained by sub-photospheric dissipation in a baryon-loaded outflow, i.e., a dissipative-photosphere scenario \citep[e.g.,][]{MeszarosRees2000,Rees2005,Peer2005,Peer2006,Beloborodov2010,Giannios2006,Giannios2012}. This interpretation is motivated by both the observations and the expected physics of dirty fireballs. Observationally, the combination of a very soft prompt spectrum and a remarkably low X-ray plateau luminosity points to a relatively low Lorentz factor, suggesting that the spectral softness may be  related to the low $\Gamma$. 
Theoretically, a lower $\Gamma$ shifts the photosphere to larger radii, making standard dissipation processes such as internal shocks more likely to occur below it and  favoring the dissipative photosphere origin for the prompt emission. A lower $\Gamma$ also implies higher comoving photon and lepton densities, which facilitate thermalization, decrease the thermalization temperature, and shift the spectral peak to lower energies.

We also examine the possible origin of the baryon-loaded jet. Two widely discussed jet-launching channels in the collapse of massive stars are neutrino-dominated accretion flows (NDAFs) and the Blandford--Znajek (BZ) mechanism. In the NDAF scenario, neutrino annihilation above the disk can power a bipolar jet, while neutrino heating simultaneously drives a baryonic wind from the disk atmosphere, naturally enhancing the mass loading of the outflow and reducing its terminal Lorentz factor \citep{Popham1999,Narayan2001,ChenBeloborodov2007,ZalameaBeloborodov2011,Metzger2008,Lei2013}. By contrast, BZ jets extract rotational energy electromagnetically from the black hole, and the polar region can remain relatively baryon poor because magnetic support and charge-separation effects help suppress baryon loading \citep{BlandfordZnajek1977,McKinney2005,Tchekhovskoy2011,Li2000}. Motivated by these considerations, we explore a picture in which EP241113a is powered by an NDAF-launched baryonic jet whose prompt emission is dominated by a dissipative photosphere. In this framework, enhanced baryon loading reduces the terminal Lorentz factor, enlarges the photospheric radius, and shifts the spectral peak into the soft X-ray band, while a sufficiently powerful jet can still account for the large radiated energy. This provides a natural explanation for an energetic yet unusually soft transient lying far from the classical Amati relation.

The paper is organized as follows. In Sect.~\ref{sec_radiation_mechanism}, we discuss the radiation mechanism of dirty fireballs, including thermalization in opaque outflows, spectral formation in the dissipative-photosphere scenario, and the application to EP241113a. In Sect.~\ref{sec_jet_launching}, we examine the jet-launching mechanism, with particular attention to NDAF-powered baryonic jets and their implications for EP241113a. Sect.~\ref{sec_summary_discussion} summarizes the main results and discussion.

\section{Radiation mechanisms in dirty fireballs}
\label{sec_radiation_mechanism}
The gamma-ray and hard X-ray emission of GRBs is generally attributed to some dissipative processes in the jet. The outflow is inferred to be unsteady, and internal shocks \textcolor{black}{above the photosphere} are one of the most widely discussed dissipative processes because they can convert bulk kinetic energy into relativistic electrons, which then radiate efficiently (e.g., via synchrotron emission) on short timescales. However, for dirty jets with low $\Gamma$, the Thomson photospheric radius can be significantly larger than the internal shock radius, as we show below \textcolor{black}{(see also \cite{Bromberg2011,Samuelsson2023})}. Consider a jet coasting with bulk Lorentz factor $\Gamma$, total isotropic equivalent luminosity $L$, and dimensionless entropy $\eta$. The Thomson photospheric radius, defined by the scattering optical depth $\tau=n_e\sigma_{\rm T}R/\Gamma = 1$, is
\begin{align}
    R_{\rm ph} &= \frac{\sigma_{\rm T} f_{\pm} L}{4 \pi \eta \Gamma^2 m_p c^3 } \notag \\
    &= 1.2\times 10^{15} f_{\pm} L_{51} \eta_{1}^{-1} \Gamma_{1}^{-2} \, \rm cm,
\end{align}
where $f_{\pm}$ accounts for electron/positron pairs per proton, while the characteristic dissipation radius is
\begin{align}
    R_{\rm dis} &\simeq 2\Gamma_{\rm s}^2 c \delta t \notag \\
    &= 6 \times 10^{12} \Gamma_{\rm s, 1}^2 \left(\frac{\delta t}{1\,{\rm s}}\right)\, \rm cm,
\end{align}
where \textcolor{black}{$\delta t$ denotes the variability timescale of the central engine, which could be much shorter than the observed variability timescale in the dissipative photosphere model 
(given that $R_{\rm dis} < R_{\rm ph}$), and $\Gamma_{\rm s}$ is the Lorentz factor of the slower shell involved in the internal collision.} 
Unless otherwise specified, we adopt the standard notation $Q_x\equiv Q/10^x$ in cgs units. 
{Therefore, for dirty fireballs with moderate Lorentz factors, internal shocks are more likely to occur below the photosphere.}
Beyond the characteristic collision radius $R_{\rm dis}$, residual internal collisions may continue to dissipate energy within the opaque outflow over a broad subphotospheric range up to $R_{\rm ph}$ \citep{Li2008}. At the large optical depths considered here, such collisions are expected to proceed through radiation-mediated shocks \citep{Bromberg2011,Levinson2012,NordinNobuoka2026}. Magnetic reconnection may also contribute to the heating, since dissipation of even a subdominant magnetic field can provide sufficient heat per photon \citep{Beloborodov2013}. The prompt emission of a dirty fireball is therefore expected to be dominated by dissipative photospheric radiation. In this case, the peak energy, $E_{\rm p}$, of the spectrum is unlikely to be set  after the jet becomes optically thin, and is more naturally regulated while the outflow is still opaque, when radiation and plasma remain strongly coupled. In this regime, repeated Compton scatterings redistribute the dissipated energy, while photon production determines whether the radiation field can maintain full thermodynamic equilibrium. Their interplay governs the spectral evolution of the expanding jet and sets the characteristic energy scale of the prompt emission \citep{Beloborodov2013,Vurm2013}.

\subsection{Peak energy in the dissipative-photosphere scenario}
At sufficiently small radii, photon production is efficient enough to maintain a Planck spectrum. The radiation temperature ($T$) is set by the comoving radiation energy density ($U_{\gamma}$),
\begin{align}
    aT^4 = U_{\gamma}
    = \frac{L_{\gamma}}{4\pi R^2 (4/3)\Gamma^2 c},
\end{align}
where $L_{\gamma}$ is the radiative luminosity, $R$ is the radius of the jet and $c$ is the speed of light.
In this fully thermalized region, the photon number density is equal to the blackbody value,
\begin{align}
\label{Eq_nph_BB}
    n_{\gamma}=n_{\rm BB}\simeq 2.4\,\frac{8\pi}{\lambda_{\rm C}^{3}}\,\Theta^{3},
\end{align}
where $\lambda_{\rm C}\equiv h/(m_{\rm e}c)$ is the Compton wavelength and $\Theta \equiv kT/(m_{\rm e}c^{2})$. 

As the jet expands, the declining particle and photon densities suppress photon production. This defines the \emph{Planck radius} $R_{\rm P}$, inside of which
radiative processes in the thermal plasma are fast enough to
enforce a Planck spectrum. Outside $R_{\rm P}$, full thermodynamic equilibrium can no longer be maintained, and the photon number becomes approximately conserved \citep{Beloborodov2013}. 
The Planck radius is determined by the condition \footnote{We assume $y \geq 1$ at the radius given by Eq.~\ref{eq_planck_ra}, as appropriate for the parameter range considered here. Otherwise, $R_{\rm P}$ is set by $y=1$ rather than by the photon-production condition, implying $R_{\rm P}=R_{\rm W}$.}
\begin{equation}
\label{eq_planck_ra}
    \dot{n}_{\gamma} t_{\rm dyn} = n_{\rm BB},
\end{equation}
where $t_{\rm dyn}=R/(\Gamma c)$ is the comoving expansion time and $\dot{n}_\gamma$ is the photon production rate. 
In the thermal plasma, the dominant photon sources are double-Compton scattering ($e+\gamma \rightarrow e+\gamma+\gamma$) and bremsstrahlung ($e+p \rightarrow e+p+\gamma$), with rates
\begin{equation}
\label{eq_dot_n_dc}
    \dot{n}_{\gamma,\rm dc} = \chi n_{\gamma} n_{\rm e} \sigma_{\rm T} c\,\Theta^{2},
\end{equation}
\begin{equation}
\label{eq_dot_n_ff}
    \dot{n}_{\gamma,\rm ff} = \xi n_{\rm e}^{2}\sigma_{\rm T} c\,\Theta^{-1/2},
\end{equation}
respectively, where the numerical factors $\chi\simeq 0.1$ and $\xi\simeq 0.06$ weakly
(logarithmically) depend on $\Theta$ \citep{Beloborodov2013}. \textcolor{black}{Here we focus on weakly magnetized baryonic outflows, for which synchrotron photon production is assumed to be subdominant. In more strongly magnetized flows, synchrotron emission from shock-heated electrons may provide an additional source of soft photons and may modify the thermalization condition \citep{Beloborodov2013, Vurm2013, Lundman2019}.} Here the comoving electron number density is
\begin{equation}
\label{Eq_n_e}
    n_{\rm e}=\frac{f_{\pm}L}{4\pi R^{2}m_{\rm p}c^{3}\Gamma\eta}.
\end{equation}


Combining Eqs.~\ref{eq_planck_ra}--\ref{eq_dot_n_ff}, we obtain simple estimates for the Planck radius in the two limiting regimes \footnote{For the fiducial parameter set adopted here, double-Compton and bremsstrahlung contribute comparably, with \(\dot{n}_{\gamma,\rm dc}/\dot{n}_{\gamma,\rm ff} \simeq \textcolor{black}{2.5}\,\epsilon_{\rm P}^{11/8}\eta_1\textcolor{black}{\Gamma_{{\rm P},0.5}^{-7/4}}f_{\pm}^{-1}L_{51}^{3/8}\textcolor{black}{R_{{\rm P},10.7}^{-3/4}}\). The resulting correction to the analytic estimate of \(kT_{\rm P}\) below is therefore only at the tens-of-percent level. Both processes are treated self-consistently in our numerical calculation (Appendix~\ref{sec_appendix_Rp_Rw}).}.
In the double-Compton-dominated case, $\dot{n}_{\gamma,\rm dc}\gg \dot{n}_{\gamma,\rm ff}$, the Planck radius is
\begin{align}
R_{\rm P}
&= 3.3\times10^{-1}
\left(
\frac{k^4\sigma_{\rm T}^2\chi^2}{\pi^3ac^{15}m_p^2m_e^4}
\eta^{-2}\Gamma_{\rm P}^{-6}f_{\pm}^{2}L_{\gamma}L^{2}
\right)^{1/4}
\notag\\
&\simeq \textcolor{black}{7.5\times10^{10}}
\epsilon_{\rm P}^{1/4}\eta_1^{-1/2}\textcolor{black}{\Gamma_{{\rm P},0.5}^{-3/2}}
f_{\pm}^{1/2}L_{51}^{3/4}\ {\rm cm}.
\end{align}
with a corresponding temperature
\begin{align}
\label{Eq_kTP_DC}
kT_{\rm P}
&= 1.1
\left(
\frac{\pi k^4c^{13}m_p^2m_e^4}{a\sigma_{\rm T}^2\chi^2}
\eta^{2}\Gamma_{\rm P}^{2}f_{\pm}^{-2}L_{\gamma}L^{-2}
\right)^{1/8}
\notag\\
&\simeq \textcolor{black}{4.0}
\epsilon_{\rm P}^{1/8}\eta_1^{1/4}\textcolor{black}{\Gamma_{{\rm P},0.5}^{1/4}}
f_{\pm}^{-1/4}L_{51}^{-1/8}\ {\rm keV}.
\end{align}
where \textcolor{black}{$\epsilon_{\rm P}\equiv L_{\gamma}(R_{\rm P})/L$} and $\Gamma_{\rm P}$ denote the radiative fraction and bulk Lorentz factor at $R_{\rm P}$, respectively. In the opposite limit where bremsstrahlung dominates, $\dot{n}_{\gamma,\rm ff}\gg \dot{n}_{\gamma,\rm dc}$, we find
\begin{align}
R_{\rm P}
&= 3.3\times10^{-2}
\left(
\frac{a^7h^{24}\sigma_{\rm T}^8m_e^4\xi^8}
{\pi^{17}c^9k^{28}m_p^{16}}
\eta^{-16}\Gamma_{\rm P}^{-10}f_{\pm}^{16}L_{\gamma}^{-7}L^{16}
\right)^{1/10}
\notag\\
&\simeq \textcolor{black}{4.4\times10^{10}}
\epsilon_{\rm P}^{-7/10}\eta_1^{-8/5}\textcolor{black}{\Gamma_{{\rm P},0.5}^{-1}}
f_{\pm}^{8/5}L_{51}^{9/10}\ {\rm cm},
\end{align}
and
\begin{align}
\label{Eq_kTP_ff}
kT_{\rm P}
&= 3.6
\left(
\frac{\pi^3k^{12}cm_p^4}
{a^3h^6\sigma_{\rm T}^2m_e\xi^2}
\eta^{4}f_{\pm}^{-4}L_{\gamma}^{3}L^{-4}
\right)^{1/5}
\notag\\
&\simeq 5.2
\epsilon_{\rm P}^{3/5}\eta_1^{4/5}
f_{\pm}^{-4/5}L_{51}^{-1/5}\ {\rm keV}.
\end{align}

These scalings show that the thermalization is more easily achieved in dirty fireballs than in classical GRBs. A lower Lorentz factor increases the comoving lepton density, boosting both the photon-production rate and the scattering depth. As a result, dirty fireballs can maintain full thermalization to a larger  $R_{\rm P}$ with a lower radiation temperature.

Beyond $R_{\rm P}$, the radiation can still remain in kinetic equilibrium with the plasma through saturated Comptonization, provided that the Compton $y$-parameter satisfies
\begin{equation}
    y \equiv 4\Theta\tau \gg 1,
\end{equation}
so that electrons and photons remain tightly coupled and share a common temperature \(T\).
The distribution still follows a Bose--Einstein distribution, but now with a positive photon chemical potential, $\mu>0$, and $n_{\gamma}<n_{\rm BB}$. This is defined as the Wien zone by \cite{Beloborodov2013}. It corresponds to a photon-starved regime: for a given radiation energy density, the radiation contains fewer photons than a blackbody, and therefore a higher mean energy per photon. 

The outer boundary of the Wien zone, namely the Wien radius $R_{\rm W}$, is defined by
\begin{equation}
\label{Eq_y}
    y=4\Theta\tau =1.
\end{equation}

Within the Wien zone, $R_{\rm P}<R<R_{\rm W}$, photon production through bremsstrahlung and double-Compton scattering becomes inefficient, so the total photon number is approximately conserved \citep{Beloborodov2013}, i.e., $N_{\gamma}(R)\simeq N_{\gamma}(R_{\rm P})$. The radiation temperature in this zone is therefore estimated from photon-number conservation together with energy conservation, $3\Gamma kT \times N_{\gamma}=\epsilon_{\gamma} E_{\rm tot}$, where $\epsilon_{\gamma}$ is the radiative efficiency and $E_{\rm tot}$ is the total fireball energy. Since the characteristic internal-collision radius is close to the Wien radius, a substantial fraction of the jet kinetic energy may  be converted to radiation at this stage. We therefore adopt $\epsilon_{\rm W}\simeq 0.5$. \textcolor{black}{Defining $\epsilon_{{\rm W}, -0.3}\equiv\epsilon_{\rm W}/10^{-0.3}$,} the Wien-zone temperature is then
\begin{align}
\label{Eq_kTW_DC}
kT_{\rm W}
&= kT_{\rm P}\,\frac{\epsilon_{\rm W}}{\epsilon_{\rm P}}\,\frac{\Gamma_{\rm P}}{\Gamma_{\rm W}} \notag\\
&\simeq \textcolor{black}{0.6}\,
\epsilon_{\rm P}^{-7/8}\textcolor{black}{\epsilon_{{\rm W}, -0.3}}
\eta_1^{1/4}\textcolor{black}{\Gamma_{{\rm P},0.5}^{5/4}}\Gamma_{{\rm W},1}^{-1}
f_{\pm}^{-1/4}L_{51}^{-1/8}\ {\rm keV},
\end{align}
if double-Compton scattering dominates photon production near \(R_{\rm P}\), and
\begin{align}
\label{Eq_kTW_ff}
kT_{\rm W}
&\simeq \textcolor{black}{0.8}\,
\epsilon_{\rm P}^{-2/5}\textcolor{black}{\epsilon_{{\rm W}, -0.3}}
\eta_1^{4/5}\textcolor{black}{\Gamma_{{\rm P},0.5}}\Gamma_{{\rm W},1}^{-1}
f_{\pm}^{-4/5}L_{51}^{-1/5}\ {\rm keV},
\end{align}
if bremsstrahlung dominates instead.

The corresponding Wien radius is determined from Eq.~\ref{Eq_y}:
\begin{align}
\label{Eq_RW_DC}
R_{\rm W}
&= 1.1
\left(
\frac{k^4\sigma_{\rm T}^{6}}
{\pi^7ac^{27}m_p^6m_e^4\chi^2}
\frac{\epsilon_{\rm W}^8}{\epsilon_{\rm P}^7}
\eta^{-6}\Gamma_{\rm P}^{10}\Gamma_{\rm W}^{-24}f_{\pm}^{6}L^{7}
\right)^{1/8}
\notag\\
&\simeq \textcolor{black}{6.0\times10^{12}}
\epsilon_{\rm P}^{-7/8}\textcolor{black}{\epsilon_{{\rm W}, -0.3}}
\eta_1^{-3/4}\textcolor{black}{\Gamma_{{\rm P},0.5}^{5/4}}\Gamma_{{\rm W},1}^{-3}
f_{\pm}^{3/4}L_{51}^{7/8}\ {\rm cm},
\end{align}
when double-Compton scattering dominates near \(R_{\rm P}\), and
\begin{align}
\label{Eq_RW_ff}
R_{\rm W}
&= 3.6
\left(
\frac{k^{12}\sigma_{\rm T}^{3}}
{\pi^2a^3c^{24}h^6m_pm_e^6\xi^2}
\frac{\epsilon_{\rm W}^5}{\epsilon_{\rm P}^2}
\eta^{-1}\Gamma_{\rm P}^{5}\Gamma_{\rm W}^{-15}f_{\pm}L^4
\right)^{1/5}
\notag\\
&\simeq \textcolor{black}{7.6\times10^{12}}
\epsilon_{\rm P}^{-2/5}\textcolor{black}{\epsilon_{{\rm W}, -0.3}}
\eta_1^{-1/5}\textcolor{black}{\Gamma_{{\rm P},0.5}}\Gamma_{\rm W,1}^{-3}
f_{\pm}^{1/5}L_{51}^{4/5}\ {\rm cm}.
\end{align}
when bremsstrahlung dominates instead.

In the Comptonization zone (\(R_{\rm W}<R<R_{\rm ph}\)), electrons and radiation are no longer in full thermal equilibrium. If dissipation persists, it can heat the electrons and keep the Compton parameter at \(y\equiv4\Theta_e\tau\sim1\), where \(\Theta_e\equiv kT_e/m_ec^2\) and \(T_e\) is the electron temperature. This unsaturated Comptonization can partly offset adiabatic cooling, broaden the spectrum, and produce a high-energy tail without substantially shifting the spectral peak \citep{Beloborodov2013,Beloborodov2017,Giannios2012}. 
\textcolor{black}{In this zone, continued heating can be supplied by collisional dissipation \citep{Beloborodov2010}, with residual internal collisions in an unsteady outflow providing a natural example \citep{Li2008}. Magnetic reconnection may also contribute if the outflow carries even a modest fraction of magnetic energy component.}
If, instead, heating above \(R_{\rm W}\) becomes weak and no substantial photon production occurs, the observed peak is further suppressed by a factor of \(\epsilon_{\rm ph}/\epsilon_{\rm W}\), as the radiative efficiency decreases from \(\epsilon_{\rm W}\) to \(\epsilon_{\rm ph}\) between \(R_{\rm W}\) and \(R_{\rm ph}\) while the photon number remains approximately conserved \citep{Vurm2013}. This gives
\begin{align}
    E_{\rm p} &\simeq 3\,\Gamma_{\rm W} kT_{\rm W} \frac{\epsilon_{\rm ph}}{\epsilon_{\rm W}}\notag \\
    &=3\Gamma_{\rm P} k T_{\rm P} \frac{\epsilon_{\rm ph}}{\epsilon_{\rm P}},
\end{align}
where $\Gamma_{\rm W}$ and $T_{\rm W}$ are the bulk Lorentz factor and comoving radiation temperature at the Wien radius. Substituting Eqs.~\ref{Eq_kTP_DC} and \ref{Eq_kTP_ff} gives
\begin{align}
\label{Eq_Ep_DC}
    E_{\rm p} &\simeq \textcolor{black}{3.7}\,
    \epsilon_{\rm P}^{-7/8}\epsilon_{{\rm ph},-1}
    \eta_1^{1/4}\textcolor{black}{\Gamma_{{\rm P},0.5}^{5/4}}
    f_{\pm}^{-1/4}L_{51}^{-1/8}\ {\rm keV},
\end{align}
for the double-Compton-dominated case, and
\begin{align}
\label{Eq_Ep_ff}
    E_{\rm p} &\simeq \textcolor{black}{5.1}\,
    \epsilon_{\rm P}^{-2/5}\epsilon_{{\rm ph},-1}
    \eta_1^{4/5}\textcolor{black}{\Gamma_{{\rm P},0.5}}
    f_{\pm}^{-4/5}L_{51}^{-1/5}\ {\rm keV},
\end{align}
for the bremsstrahlung-dominated case. We consider $\epsilon_{\rm ph} \ll \epsilon_{\rm P}$, motivated by the physical considerations for our dirty fireball case: At $R_{\rm P}\sim 10^{11}\,$cm, the flow remains deep in the opaque region and may not yet have completed its acceleration, as expected for a compact Wolf--Rayet progenitor in the collapsar picture \citep{Drenkhahn2002,Beloborodov2013,Beloborodov2017,Vurm2016}. The radiation field may therefore still carry a substantial fraction of the jet power, suggesting $\epsilon_{\rm P}\sim 1$ \citep{Beloborodov2017,Vurm2016}. By contrast, at the much larger photospheric radius, $R_{\rm ph}\sim 10^{14}-10^{15}\,\rm cm$, a substantial fraction of the dissipated energy may be converted back into bulk kinetic energy rather than radiated away. Although further dissipation, such as residual collisions, may still occur, it is expected to be much less efficient \citep{Li2008}. Accordingly, $\epsilon_{\rm ph}$ is expected to be significantly smaller than  $\epsilon_{\rm P}$.

These expressions show that the lower Lorentz factor of dirty fireballs leads to a larger Wien radius and a softer spectral peak. For representative dirty-fireball parameters, $\Gamma_{\rm P}\sim {\rm a\ few}$ and $\eta \sim 10$, the Wien radius can be up to \(\sim 10^{3}\) times larger than in classical GRBs with otherwise similar parameters. 
Correspondingly, if we compare these dirty-fireball parameters with representative classic GRB values, $\eta\simeq100$ and $\Gamma_{\rm P}\simeq30$, Eqs.~\ref{Eq_Ep_DC} and \ref{Eq_Ep_ff} imply that the predicted $E_{\rm p}$ is reduced to  a few percent to $\lesssim20\%$ of that of classic GRBs. The degree of the reduction depends on  the fireball parameters and on whether the photon production near $R_{\rm P}$ is dominated by double-Compton scattering or bremsstrahlung. Dirty fireballs can thus remain as energetic as classic GRBs while exhibiting much softer prompt emission spectra.

\subsection{The case of EP241113a}

\begin{figure}[htbp]
\centering
\includegraphics[width=0.9\linewidth]{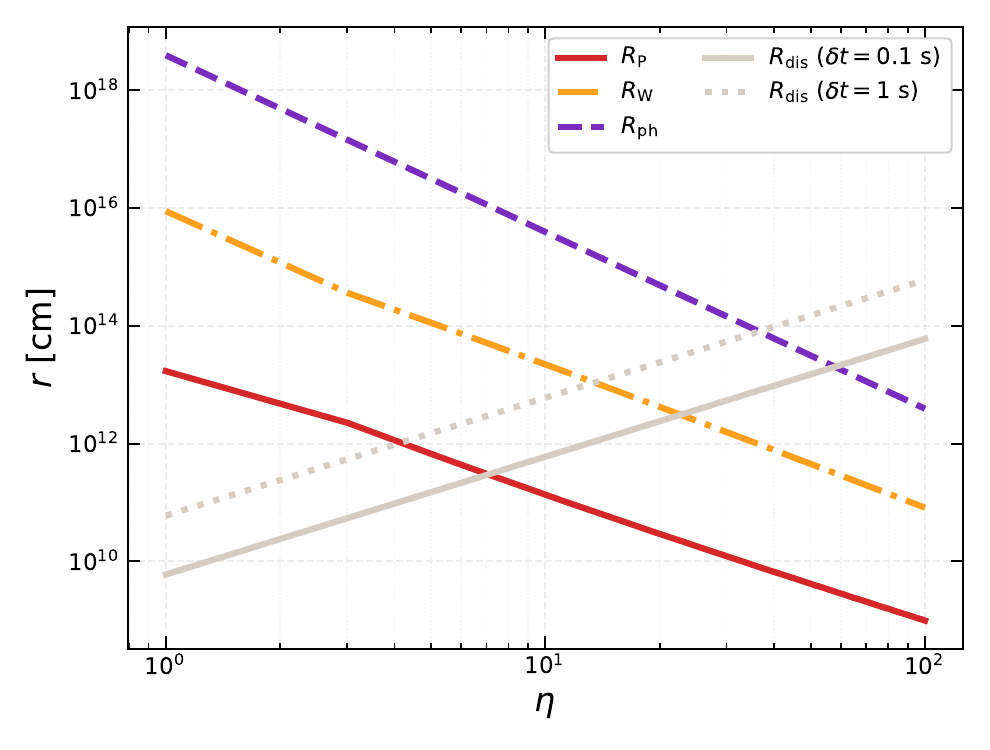} 
\caption{\textbf{Characteristic radii of a dirty fireball as a function of the dimensionless entropy $\eta$.} Shown are the Planck radius $R_{\rm P}$ (red solid), the Wien radius $R_{\rm W}$ (orange dot--dashed), the Thomson photospheric radius $R_{\rm ph}$ (purple dashed), and the dissipation radius $R_{\rm dis}\sim 2\Gamma_{\rm W}^{2}c\delta t$ for $\delta t=0.1$~s (gray solid) and $1$~s (gray dotted). The fiducial parameters are $f_{\pm}=1$, $\epsilon_{\rm P}=1$, $\epsilon_{\rm W}=0.5$, $\Gamma_{\rm P}=\max(\eta/2,1)$, and $\Gamma_{\rm W}=\eta$. The calculation adopts $\eta_\gamma=0.03$, a fixed X-ray luminosity $L_{\gamma, \rm iso}=10^{50}\ {\rm erg\ s^{-1}}$, and a total jet luminosity $L=L_{\gamma, \rm iso}/\eta_\gamma$.}
\label{fig_radius}
\end{figure}

We now apply this framework to EP241113a. To assess whether its prompt emission can be interpreted in the dissipative-photosphere scenario, we compare in Fig.~\ref{fig_radius} the Thomson photospheric radius $R_{\rm ph}$, the dissipation radius $R_{\rm dis}$, the Planck radius $R_{\rm P}$, and the Wien radius $R_{\rm W}$. The WXT data of EP241113a show a peak X-ray luminosity of $L_X \simeq 5\times10^{49}\ {\rm erg\ s^{-1}}$ in the $0.5$--$4\ {\rm keV}$ band and an extremely soft spectrum (photon index $\Gamma_X>2$ at 95\% confidence), implying that most of the emission emerges in X-rays \citep{Dai2026}. 
The observed minimum variability timescale is not well constrained. However, the duration of an individual pulse is limited to $\lesssim 30/(1+z)\,{\rm s}$ with $z=1.53$ \citep{Dai2026}, suggesting that the minimum variability timescale should be smaller than $\sim 10\,{\rm s}$. The central engine variability timescale could be much shorter, given the compact size of the central engine. We therefore consider two representative values for the central engine variability timescale, $\delta t=0.1\,{\rm s}$ and $1\,{\rm s}$, in Fig.~\ref{fig_radius}. \textcolor{black}{For the fiducial model shown in Fig.~\ref{fig_radius}, we adopt $f_{\pm}=1$, $\epsilon_{\rm P}=1$, $\epsilon_{\rm W}=0.5$, $\Gamma_{\rm P}=\max(\eta/2,1)$, and $\Gamma_{\rm W}=\eta$. The afterglow modeling in \citet{Dai2026} gives an inferred kinetic energy of $E_{\rm k}\simeq 10^{52}-10^{53}\,{\rm erg}$. Comparing this with $E_{\gamma,\rm iso}\simeq 10^{51}\,{\rm erg}$ suggests a radiative efficiency of $\eta_\gamma\sim0.01$--$0.1$ for EP241113a. We therefore adopt $\eta_\gamma=0.03$ as the fiducial value and set the total jet luminosity to $L=L_{\gamma,\rm iso}/\eta_\gamma$, with $L_{\gamma,\rm iso}\sim10^{50}\,{\rm erg\,s^{-1}}$.}
\textcolor{black}{We find that for $\eta\sim20$, as inferred for EP241113a, the dissipation radius lies close to the Wien radius and remains well below the Thomson photosphere, $R_{\rm dis}\sim R_{\rm W}\ll R_{\rm ph}$. The corresponding photospheric angular timescale is $t_{\rm ph}\sim R_{\rm ph}/(2\Gamma^2c)\sim 20\,{\rm s}\,(\Gamma/20)^{-5}$, broadly consistent with the observed variability timescale. 
}

\begin{figure}[htbp]
\centering
\includegraphics[width=0.9\linewidth]{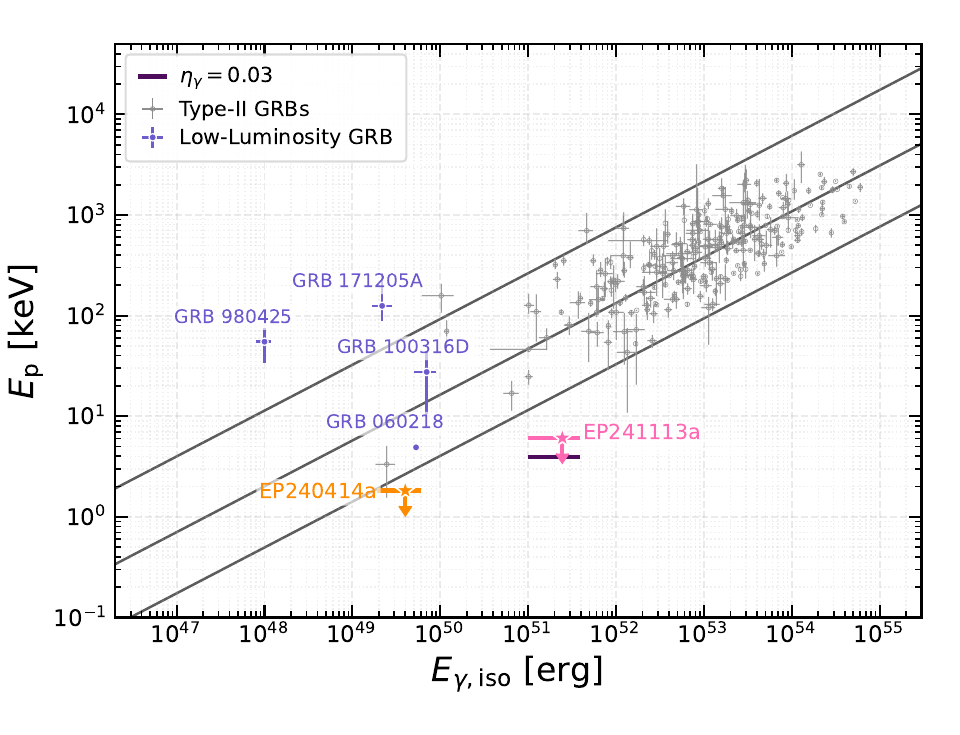}
\caption{\textbf{\texorpdfstring{$E_{\rm p}$--$E_{\gamma,\rm iso}$}{Ep--Eiso} plane.} The black horizontal line shows the position expected for EP241113a in the dissipative-photosphere scenario. The calculation adopts $f_{\pm}=1$, $\epsilon_{\rm P}=1$, $\epsilon_{\rm W}=0.5$, $\eta=20$, $\Gamma_{\rm P} = \eta / 2$, $\Gamma_{\rm W}=\eta$, and a fixed X-ray luminosity $L_{\gamma, \rm iso}=10^{50}\ {\rm erg\ s^{-1}}$. The total jet luminosity is taken to be $L=L_{\gamma, \rm iso}/\eta_{\gamma}$, with $\eta_{\gamma} = \epsilon_{\rm ph} = 0.03$.}
\label{fig_Ep}
\end{figure}

We further examine whether this scenario can reproduce the observed soft spectral peak of EP241113a. \textcolor{black}{Adopting the fiducial parameter set described above, together with $\Gamma_{\rm W} \simeq \eta=20$  and $\Gamma_{\rm P}=\eta/2$, we numerically compute the peak energy (see Appendix~\ref{sec_appendix_Rp_Rw}). The resulting spectral peak is shown by the horizontal line in Fig.~\ref{fig_Ep}. We find that the location of EP241113a can be reproduced for $\Gamma_{\rm P} \lesssim \eta/2 $.} 
Such a value of $\Gamma_{\rm P}$ is physically reasonable. The corresponding Planck radius is only a few $\times 10^{10}\,$cm, where the outflow may still be radiation dominated and not yet fully accelerated. 
By contrast, the Wien radius is much larger, reaching $\sim 10^{12}$--$10^{13}\, \rm cm$, where a substantial fraction of the fireball energy may have already been converted into bulk kinetic energy and the Lorentz factor is expected to be close to its terminal value.

EP241113a can therefore be naturally interpreted as a dirty-fireball event radiating through a dissipative photosphere. In this picture, the relatively low Lorentz factor significantly suppresses the spectral peak energy, as implied by Eqs.~\ref{Eq_Ep_DC} and \ref{Eq_Ep_ff}. The dissipative-photosphere scenario thus provides a natural explanation for why EP241113a occupies the energetic yet low-$E_{\rm p}$ region of the $E_{\rm p}$--$E_{\gamma,\rm iso}$ plane.

\section{Jet-Launching Mechanisms for Dirty Fireballs}
\label{sec_jet_launching}


{The deviation of EP241113a from the Amati relation may indicate that the jet-launching engine operates differently from those in classical GRBs and X-ray flashes that follow this relation. In the dirty-fireball interpretation, such an engine would produce jets with lower Lorentz factors, $\Gamma$, than those of GRBs with comparable energetics.}

{The most popular model for the GRB central engine invokes a stellar-mass black hole surrounded by a neutrino-dominated accretion flow at an extremely high accretion rate. In such a system, the jet may be powered either by neutrino-antineutrino annihilation above the disk or by extraction of the black-hole spin energy through the Blandford--Znajek mechanism \citep{BlandfordZnajek1977}. Both channels have been extensively investigated in previous studies \citep[e.g.,][]{Popham1999,Lee2000,Narayan2001,DiMatteo2002,Liu2007,Liu2017,Janiuk2007,Lei2013}.}

Motivated by the higher neutrino-driven baryon loading expected in NDAF engines \citep{Lei2013}, we speculate that EP241113a could be powered by an NDAF-launched baryonic jet.
In the NDAF regime, bounded by the neutrino-ignition and neutrino-trapping accretion rates $\dot{M}_{\rm ign}$ and $\dot{M}_{\rm trap}$, a fraction of neutrinos annihilate above the disk and power a jet via $\nu\bar{\nu}\rightarrow e^+e^-$. The annihilation power is \citep{ZalameaBeloborodov2011}
\begin{equation}
\label{Eq_dot_Evv}
\dot{E}_{\nu\bar{\nu}} \simeq 1.1 \times 10^{52}
\left( \frac{R_{\rm ms}}{2} \right)^{-4.8}
\left( \frac{m_{\rm BH}}{3} \right)^{-3/2}
\dot{m}_{\rm acc}^{9/4}\ {\rm erg\ s^{-1}},
\end{equation}
where $m_{\rm BH}\equiv M_{\rm BH}/M_\odot$, $\dot{m}_{\rm acc}\equiv \dot{M}_{\rm acc}/(M_\odot\,{\rm s^{-1}})$, and $R_{\rm ms}\equiv r_{\rm ms}/r_g$ with $r_g\equiv GM_{\rm BH}/c^2$.

Neutrino heating above the disk surface drives a wind that loads baryons into the polar region. The jet mass-loading rate is \citep{Metzger2008,Qian1996}
\begin{align}
\label{Eq_dot_Mjvv}
\dot{M}_{j,\nu\bar{\nu}} &= \dot{M}_{\nu}\,\theta_{\nu\bar{\nu}}^{2}/2 \notag \\
&= 7.0\times10^{-7}\,A^{0.85} B^{-1.35} C^{0.22}\,\theta_{\nu\bar{\nu},-1}^{2}\,\alpha_{-1}^{0.57}\,\epsilon_{\nu, -1}^{1.7} \notag \\
&\quad \times \left(\frac{R_{\rm ms}}{2}\right)^{0.32}\,\dot{m}_{{\rm acc}, -1}^{1.7}\left(\frac{m_{\rm BH}}{3}\right)^{-0.9}\left(\frac{\xi_{\rm d}}{2}\right)^{0.32}\,M_{\odot}\,{\rm s}^{-1},
\end{align}
where $\dot{M}_{\nu}$ is the isotropic mass-loss rate  and $\theta_{\nu\bar{\nu}}$ is the jet half-opening angle; $A$, $B$, and $C$ are the Kerr-disk relativistic correction factors, $\alpha$ is the viscosity parameter, $\epsilon_{\nu}$ is the neutrino emission efficiency, and $\xi_{\rm d}$ is the disk radius in units of $r_{\rm ms}$ \citep{Lei2013}.

The BH mass and spin evolve as
\begin{align}
&\frac{dM_{\rm BH} c^2}{dt}=\dot{M}_{\rm acc}c^2E_{\rm ms}, \notag \\
&\frac{da_{\rm BH}}{dt}=\frac{\dot{M}_{\rm acc}L_{\rm ms}c}{GM_{\rm BH}^2}-\frac{2a_{\rm BH}\dot{M}_{\rm acc}E_{\rm ms}}{M_{\rm BH}},
\end{align}
where $E_{\rm ms}$ and $L_{\rm ms}$ are the innermost stable circular orbit (ISCO) specific energy and angular momentum, and $a_{\rm BH}$ is the dimensionless BH spin. We define the time-averaged dimensionless entropy, namely the maximum energy per baryon of a thermally accelerated fireball, as
\begin{equation}
\label{eq_eta}
\bar{\eta}=\frac{\int\left(\dot{E}_{\nu\bar{\nu}}+\dot{M}_{j,\nu\bar{\nu}}c^2\right)dt}{\int \dot{M}_{j,\nu\bar{\nu}}c^2\,dt}.
\end{equation}

The jet power (Eq.~\ref{Eq_dot_Evv}) and mass-loading rate (Eq.~\ref{Eq_dot_Mjvv}) depend sensitively on the black-hole spin, $a_{\rm BH}$, and the dimensionless accretion rate, $\dot{m}_{\rm acc}$. The resulting baryon loading of the outflow (Eq.~\ref{eq_eta}) is therefore also strongly regulated by these two parameters. We thus explore a parameter grid spanning $a_{\rm BH,0}\in(0.1,\,0.998)$ and $\dot{m}_{\rm acc}\in(0.01,\,1)$, adopting an accretion duration of $\sim 100\,{\rm s}$ and fixing the remaining quantities to the fiducial values of \cite{Lei2013}.

\begin{figure}[htbp]
    \centering
    \includegraphics[width = 1.0\linewidth]{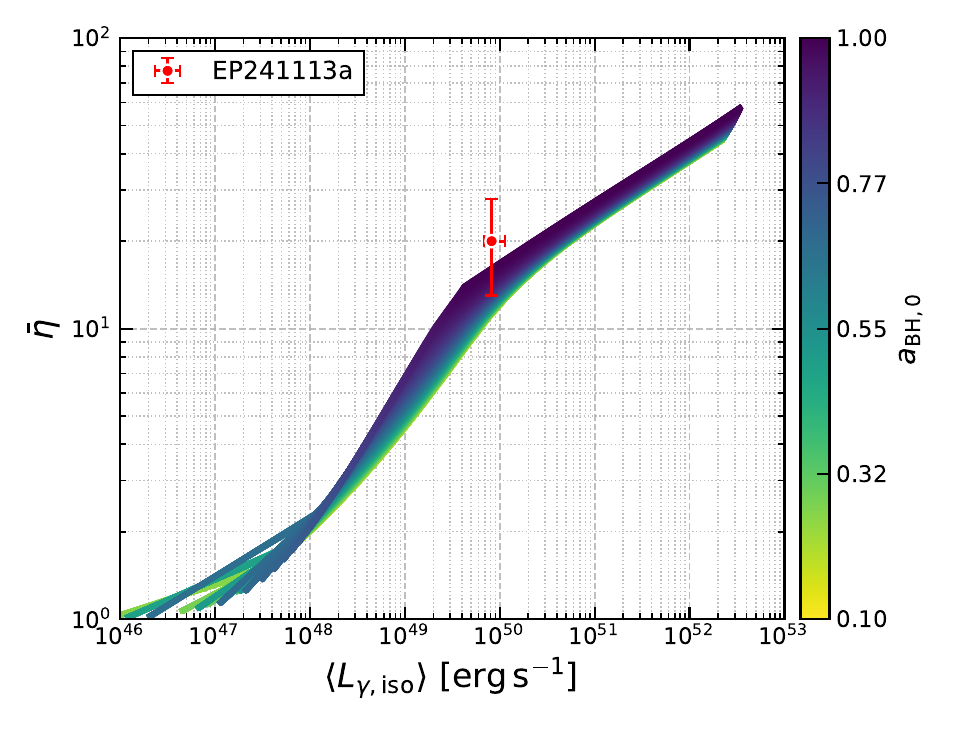}
    \caption{\textbf{Dimensionless entropies predicted for NDAF-powered jets.} Curves show the dimensionless entropy $\bar{\eta}$ versus the time-averaged isotropic-equivalent prompt luminosity $\langle L_{\gamma,\rm iso}\rangle$. The color represents the initial black-hole spin $a_{\rm BH,0}$, while each curve is traced by varying $\dot{m}_{\rm acc}=0.01$--1 with the other parameters fixed at the fiducial values of \cite{Lei2013}. The red point with error bars marks EP241113a \citep{Dai2026}.}
    \label{fig_Gamma}
\end{figure}

For each parameter set, we compute the jet luminosity and the corresponding Lorentz factor following the above prescription, and convert the jet power into the isotropic-equivalent prompt luminosity \textcolor{black}{using a fiducial radiative efficiency $\eta_\gamma=0.03$}. Fig.~\ref{fig_Gamma} shows the predicted $\eta$ of NDAF-launched jets. The curves are color-coded by the initial spin $a_{\rm BH,0}$. Along each curve, both $\eta$ and $L_{\gamma,\rm iso}$ increase with $\dot{m}_{\rm acc}$.

Overall, NDAF engines tend to produce lower Lorentz factors than those of classical GRBs (typically $\Gamma\gtrsim 100$). For $\eta_\gamma=0.03$, the predicted jets have $\eta\sim$ a few at $L_{\gamma,\rm iso}\lesssim 10^{50}\,{\rm erg\,s^{-1}}$ and approach $\eta\sim 60$ at the highest luminosities, $L_{\gamma,\rm iso}\sim 10^{52}\,{\rm erg\,s^{-1}}$. Such modest Lorentz factors naturally accommodate dirty-fireball transients and are broadly consistent with the low-$\Gamma$ requirement for EP241113a (see Fig.~\ref{fig_Gamma}), supporting an NDAF central engine in this event. {Meanwhile, the power of the BZ mechanism depends sensitively on the magnetic flux threading  the BH horizon.  If the magnetic field is weak, the BZ power could be correspondingly reduced, and the neutrino-annihilation power may become dominant.}

\begin{figure}[htbp]
\centering
\includegraphics[width = 1.0\linewidth]{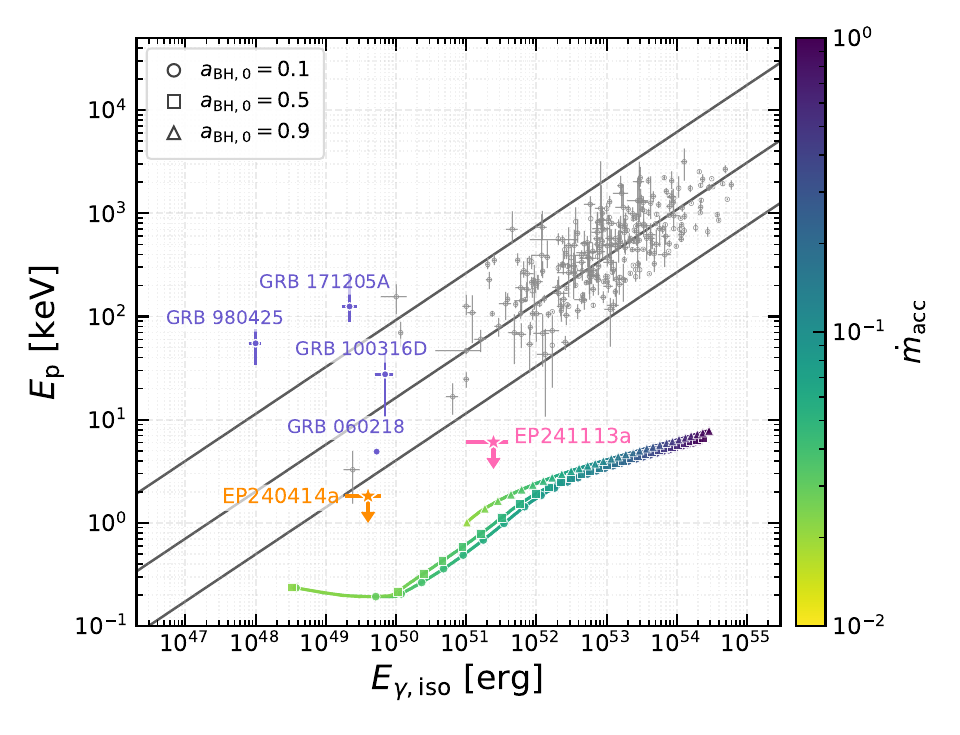}
\caption{\textbf{\texorpdfstring{$E_{\rm p}$--$E_{\gamma,\rm iso}$}{Ep--Eiso} plane for NDAF $\nu\bar{\nu}$-annihilation jets.} The colored sequences show the model tracks of $\nu\bar{\nu}$-annihilation-powered baryonic jets in the dissipative-photosphere scenario. The different marker shapes indicate different initial black-hole spin $a_{\rm BH,0}$, while the color bar represents the accretion rate $\dot{m}_{\rm acc}$. The calculation adopts $\eta_\gamma=\epsilon_{\rm ph}=0.03$, $f_{\pm}=1$, $\epsilon_{\rm P}=1$, $\epsilon_{\rm W}=0.5$, $\Gamma_{\rm P}=\max(\eta/2,1)$, and $\Gamma_{\rm W}=\eta$.}
\label{fig_Ep_Eiso_NDAF}
\end{figure}

Within the NDAF framework, we next evaluate the prompt spectral peak energy in the dissipative-photosphere scenario. 
We adopt $f_{\pm}=1$, $\epsilon_{\rm P}=1$, $\epsilon_{\rm W}=0.5$, $\Gamma_{\rm P}=\max(\eta/2,1)$, $\Gamma_{\rm W}=\eta$, and $\eta_\gamma=\epsilon_{\rm ph}=0.03$. We take the jet luminosity and dimensionless entropy from Fig.~\ref{fig_Gamma}. For these parameters, our numerical calculation (Appendix~\ref{sec_appendix_Rp_Rw}) yields a Planck radius of $\sim 10^{10}$--$10^{11}\,\rm cm$ and a Wien radius of $\sim 10^{13}$--$10^{14}\,\rm cm$, with $R_{\rm P} \ll R_{\rm W} \ll R_{\rm ph}$. The resulting $E_{\rm p}$--$E_{\gamma,\rm iso}$ relation is shown in Fig.~\ref{fig_Ep_Eiso_NDAF}. We find that dirty NDAF jets occupy a distinct region in the $E_{\rm p}$--$E_{\gamma,\rm iso}$ plane, lying systematically below the observed Amati relation, with lower $E_{\rm p}$ at a given $E_{\gamma,\rm iso}$. In the dissipative-photosphere scenario, this offset arises naturally from their lower Lorentz factors, which shift the spectral peak to lower energies, below  those of X-ray flashes with comparable isotropic-equivalent energies. Consequently, NDAF-launched dirty jets can naturally account for energetic yet much softer transients (such as EP241113a) relative to the standard Amati relation.

\section{Summary and discussion}
\label{sec_summary_discussion}

{EP241113a is an unusual fast X-ray transient: its isotropic-equivalent radiation energy is comparable to that of classical GRBs, yet its prompt spectrum peaks in the soft X-ray band, with a low $E_{\rm p}$ and no contemporaneous $\gamma$-ray detection. Its afterglow is also unusual, featuring an exceptionally faint X-ray plateau. The temporal behavior of the plateau is consistent with that expected from a jet freely expanding into a wind medium. Under this interpretation, the low plateau luminosity implies a low jet Lorentz factor, $\Gamma \sim 20$ \citep{Dai2026}. EP241113a is therefore consistent with an energetic ``dirty fireball.''}

{In this paper, we have shown that the photosphere shifts to larger radii in low $\Gamma$ jets, making standard dissipation processes such as internal shocks more likely to occur below it. As a result, the prompt emission naturally arises from a dissipative photosphere. For a given luminosity, a lower bulk Lorentz factor also implies higher comoving photon and lepton densities, which enhance soft photon production and increase the scattering optical depth, thereby facilitating thermalization. The resulting lower temperature at a larger thermalization radius then leads to a lower spectral peak energy, $E_{\rm p}$. This picture provides a natural explanation for the low $E_{\rm p}$ of EP241113a despite the fact that its isotropic equivalent radiation energy is comparable to that of classical GRBs.}

\textcolor{black}{Our calculation applies to a relatively steady outflow. In an unsteady jet, internal collisions may introduce local variations in the density, photon number, and dissipation rate \citep{NordinNobuoka2026}. The radii $R_{\rm dis}$, $R_{\rm W}$, and $R_{\rm ph}$ should then be regarded as characteristic values for the portion of the jet that dominates the emission. Since the dissipation of the dirty fireball considered here occurs well below the photosphere, such inhomogeneities can be partly smoothed out before photon decoupling. This deep subphotospheric dissipation may also be consistent with the relatively low radiative efficiency, $\eta_\gamma\sim0.03$.}

Beyond the radiation mechanism, we further investigate the possible origin of such baryon-rich outflows. Motivated by the expectation that neutrino-driven baryon loading in NDAF-powered jets can be larger than that in BZ jets \citep{Lei2013}, we compute the jet luminosity and Lorentz factor within the NDAF framework. The resulting jets are substantially dirtier than classical GRB jets: $\eta \sim {\rm few}$ at $L_{\gamma,\rm iso}\lesssim 10^{50}\,{\rm erg\,s^{-1}}$, reaching $\eta \sim 60$ only at the highest luminosities, $L_{\gamma,\rm iso}\sim 10^{52}\,{\rm erg\,s^{-1}}$.
Using these NDAF predictions as inputs to the dissipative-photosphere model, we then calculate the corresponding $E_{\rm p}$ and synthesize the $E_{\rm p}$--$E_{\gamma,\rm iso}$ distribution. The model produces a population that is systematically softer than the canonical Amati relation at comparable energies. Consequently, NDAF-launched dirty jets could account for energetic yet much softer transients such as EP241113a. 

\textcolor{black}{We note that the high-\(\dot{m}_{\rm acc}\) end of the NDAF tracks in Fig.~\ref{fig_Ep_Eiso_NDAF} is relatively insensitive to the initial black-hole spin and corresponds to luminous transients with relatively high \(E_{\rm p}\). However, the predicted spectral peaks still lie mainly in the soft X-ray band. The observed sample of such events remains limited, which is due to lack of redshift measurements for a large fraction of EP events.   Thus, while the apparent lack of a large observed population of such events may suggest that such high-accretion-rate NDAF phases are rare or short-lived, a larger WXT sample with redshift measurements is needed to test this implication.}

More generally, the deviation of EP241113a from the Amati relation may reflect a transition from a non-thermal-dominated regime to the photosphere-dominated regime. GRBs that obey the Amati relation may be dominated by a non-thermal component,  probably of the synchrotron origin \citep{Zhang2002, Zhang2011}, as expected in relatively clean, significantly magnetized BZ jets in which a bright photospheric component is suppressed \citep{Gao2015}.  By contrast, EP241113a is produced by a much dirtier, low-$\Gamma$ jet \citep{Dai2026}, likely launched by a weakly magnetized, neutrino-annihilation-dominated engine.

\begin{figure}[htbp]
\centering
\includegraphics[width = 0.9\linewidth]{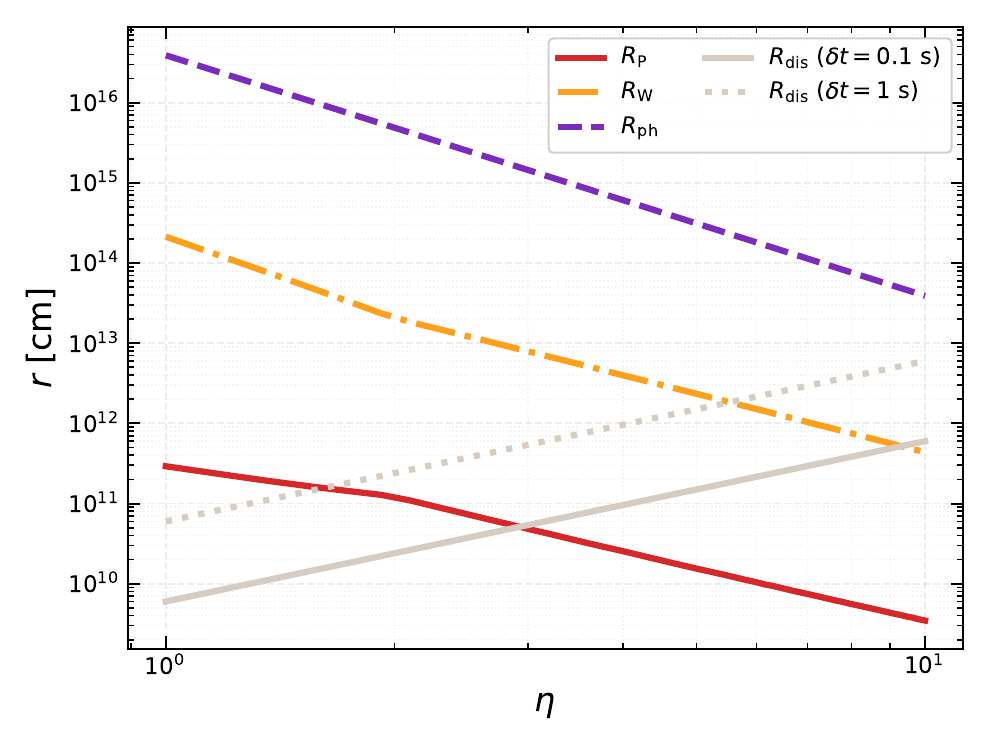}
\caption{{\bf Characteristic radii of a dirty fireball as a function of the dimensionless entropy $\eta$, assuming $L_{\gamma, \rm iso}=10^{48} \, {\rm erg\, s^{-1}}$, as relevant to EP240414a.} The line styles have the same meanings as in Fig.~\ref{fig_radius}, and the same fiducial parameters are adopted except for the luminosity.}
\label{fig_radius_EP240414a}
\end{figure}

Another extragalactic fast X-ray transient, EP240414a, also appears as an outlier to the Amati relation \citep{Sun2025}, with an isotropic energy of $E_{\gamma,\rm iso}\sim E_{X,\rm iso}\simeq 5\times10^{49}\,{\rm erg}$, where $E_{X,\rm iso}$ is the isotropic-equivalent X-ray energy, substantially lower than that of classical GRBs. This offset can likewise be understood in the dissipative-photosphere framework if the jet Lorentz factor is only a few. As shown in Fig.~\ref{fig_radius_EP240414a}, for $\eta \sim$ a few, the dissipation radius is expected to lie near $R_{\rm W}$ and remain deeply embedded within the photosphere. In this low-$\Gamma$ regime, the model predicts a soft X-ray spectral peak at several tenths of a keV, consistent with the observed upper limit for EP240414a, $E_{\rm p}<1.8\,{\rm keV}$. Within the NDAF framework, such a low Lorentz factor is also achievable, but only for relatively low initial black-hole spin and accretion rate. A weaker central engine then produces an isotropic luminosity of only $L_{\gamma, \rm iso} \sim 10^{48} \, \rm erg \, s^{-1}$, as shown in Fig.~\ref{fig_Gamma} and Fig.~\ref{fig_Ep_Eiso_NDAF}. This is in good agreement with the observed luminosity of EP240414a. EP240414a may therefore also originate from photospheric emission from an NDAF-powered dirty jet, whose progenitor would then likely have had a relatively low initial black-hole spin and accretion rate.

\begin{acknowledgments}
We would like to thank Wei-Hua Lei, Tong Liu and Jian-He Zheng  for
valuable discussions. This work is supported by the
National Natural Science Foundation of China (grant
Nos. 12333006 and  12121003) and  the
Fundamental Research Funds for the Central Universities (KG202502). We are grateful to the High Performance Computing Center of
Nanjing University for doing the numerical calculations in this
paper on its blade cluster system.
\end{acknowledgments}

\clearpage

\appendix

\section{Numerical Calculation of the Planck and Wien Radii}
\label{sec_appendix_Rp_Rw}
For the numerical calculation, we first solve Eq.~\ref{eq_planck_ra} for the maximum Planck radius, $R_{\rm P,max}$, and evaluate the Compton parameter $y$ there. If $y\leq 1$, the Planck and Wien radii coincide, $R_{\rm P}=R_{\rm W}$, and are set by the condition $y=1$. If $y>1$, then $R_{\rm P}=R_{\rm P,max}$, as in the parameter range considered in this work, and we determine $R_{\rm W}$ by evolving the photon-to-baryon ratio, $f(r)\equiv n_\gamma/n_{\rm b}$, where $n_{\rm b}$ is the comoving baryon number density, rather than using the analytic freeze-out approximation at $R_{\rm P}$.

Using $dt=dr/(\Gamma c)$ for a steady relativistic outflow, $f$ evolves as \citep{Beloborodov2013}
\begin{equation}
\frac{df}{dr}
=
\frac{\dot n_\gamma}{n_{\rm b}\Gamma c}
=
\frac{\dot n_{\rm ff}+\dot n_{\rm DC}}{n_{\rm b}\Gamma c}
\left(1-\frac{n_\gamma}{n_{\rm BB}}\right),
\label{eq:f_evolution}
\end{equation}
where $\dot n_{\rm ff}$ and $\dot n_{\rm DC}$ are the bremsstrahlung and double-Compton photon production rates, and $n_{\rm BB}(\Theta)$ is the Planck photon density. The factor $(1-n_\gamma/n_{\rm BB})$ accounts for absorption and ensures zero net production in thermal equilibrium. Together with the relation $\Theta=\Theta(f)$ derived above, Eq.~\ref{eq:f_evolution} gives a closed equation for $f(r)$.

\clearpage

\bibliography{EP241113a-Amati}{}
\bibliographystyle{aasjournalv7}

\end{document}